\documentclass[pra,aps,showpacs,twocolumn]{revtex4-1}
\usepackage{amsmath}
\usepackage{amssymb}
\usepackage{graphicx}
\newcommand{\ket}[1]{|#1\rangle}
\newcommand{\bra}[1]{\langle #1|}

\usepackage{color,soul}
\setstcolor{blue}
\setulcolor{red}

\begin{document}
\title{Theorem on the existence of a nonzero energy gap in adiabatic quantum computation}
\author{Da-Jian Zhang}
\affiliation{Department of Physics, Shandong University, Jinan 250100, China}
\author{Xiao-Dong Yu}
\affiliation{Department of Physics, Shandong University, Jinan 250100, China}
\author{D. M. Tong}
\email{tdm@sdu.edu.cn}
\affiliation{Department of Physics, Shandong University, Jinan 250100, China}
\pacs{03.67.Ac,02.10.Ox,03.67.Lx}
\date{\today}
\begin{abstract}
Adiabatic quantum computation, based on the adiabatic theorem, is a
promising alternative to conventional quantum computation. The
validity of an adiabatic algorithm depends on the existence of a
nonzero energy gap between the ground and excited states.
However, it is difficult to ascertain the exact value of the energy
gap. In this paper, we put forward a theorem on the
existence of nonzero energy gap for the Hamiltonians used in
adiabatic quantum computation. It can help to effectively identify a
large class of the Hamiltonians  without energy-level crossing between the ground and excited states.

\end{abstract}
\maketitle

Adiabatic quantum computation \cite{E. Farhi} offers a promising
alternative to the conventional quantum computation \cite{D. Deutsch,D. P. Divincenzo}.
It is polynomially equivalent to the circuit model in computational power \cite{W. van Dam,D. Aharonov,A. Mizel}, and may be more physically realistic and implementable than the circuit model due to its inherent robustness against some types of errors \cite{Childs,J. Roland,J.Aberg}.
A number of adiabatic quantum algorithms have been proposed for solving various problems \cite{E. Farhi,T. Hogg,F. Gaitan,Gaitan2,Childs2,Ralf,Hofmann,Schaller,Roland,W. van Dam,S. Garnerone,Xinhua Peng,Amin}, and several schemes have been experimentally demonstrated \cite{Steffen,Xinhua Peng,R.Harris,Zheng,Johnson}. Development of adiabatic quantum computation continues to be of great interest.

An adiabatic quantum algorithm is performed by a quantum system with a time-dependent Hamiltonian, which may be generally expressed as
\begin{equation}
  H(s) = (1-s) H_i + s H_p,
  \label{eq:H(s)}
\end{equation}
where $s=\frac{t}{T}$, for $t\in[0,T]$, is a time-dependent parameter,
$H_i$ is the initial Hamiltonian whose ground state is easy to prepare,
and $H_p$ is the problem Hamiltonian whose ground state encodes the
solution to a problem. In the practical applications, $H_p$ is usually taken as a diagonal matrix in the computational basis,
\begin{eqnarray}
H_p= \sum_z f_z\ket{z}\bra{z}, \label{problem H1}
\end{eqnarray}
where $f_z$ are real numbers, and $z\in\{0,1\}^n$ in the $n$-bit instance. The time-dependent Hamiltonian interpolates
smoothly from the initial Hamiltonian to the problem Hamiltonian.
The adiabatic theorem indicates that the final state of the quantum
system, starting from the ground state of the initial Hamiltonian, is close to the ground state of the problem Hamiltonian if
the time-dependent Hamiltonian varies sufficiently slowly.
With an appropriate measurement on the system, solutions of the problem are yielded with high probability.

\begin{figure}[htbp]
\begin{center}
\includegraphics[width=.25\textwidth]{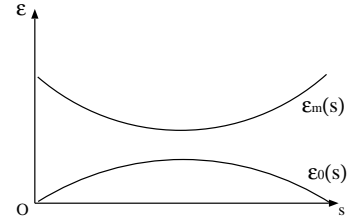}
\end{center}
\caption{A sketch of a spectrum with a nonzero energy gap.}
\label{fig.1}
\end{figure}

\begin{figure}[htbp]
\begin{center}
\includegraphics[width=.25\textwidth]{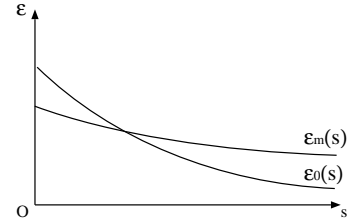}
\end{center}
\caption{A sketch of a spectrum with an energy-level crossing.}
\label{fig.2}
\end{figure}

Adiabatic quantum computation is based on the adiabatic evolution. The ``slowness" required by the adiabatic theorem is usually encoded
in the adiabatic condition,  $\frac{|\bra{\psi_m(s)}\dot
H(s)\ket{\psi_0(s)}|}{T(\Delta\varepsilon_m(s))^2}\ll 1$,
where $\Delta\varepsilon_m(s)=\varepsilon_m(s)-\varepsilon_0(s)$, $m\neq 0$, is  the energy gap between the ground state $\ket{\psi_0(s)}$ and the $m$th excited state $\ket{\psi_m(s)}$ of $H(s)$ \cite{M. Born,Bohm,Messiah}.
It shows that the energy gap $\Delta\varepsilon_m(s)$ plays an
important role in adiabatic quantum computation. The validity of an
adiabatic algorithm, i.e., the existence of a finite runtime $T$,
completely depends on the existence of a nonzero gap, while
the efficiency of the algorithm, i.e., the scaling of the runtime, depends on the value of the nonzero gap. An adiabatic
quantum algorithm works only if a nonzero energy gap always
exists during the evolution (see Fig. \ref{fig.1}), i.e., $\Delta\varepsilon_m(s)>0$
for $s\in[0,1)$ \footnote{A level crossing at the end of the runtime does not affect the validity of the algorithm, since each ground state of $H_p$ encodes a solution to the problem.}. The quantum system would fail to keep at the instantaneous ground state of $H(s)$ if the energy-level crossing between the ground and excited states happens during the evolution time (see Fig. \ref{fig.2}).
However, it is quite difficult to ascertain the
exact value of the energy gap for the Hamiltonians used in adiabatic
quantum computation. Due to the difficulty,  researchers have to resort to numerical simulations to evaluate the runtime of the adiabatic algorithm.  For example, $3925$ instances were calculated to
simulate the adiabatic algorithm for solving the $3$-SAT problem in
Ref. \cite{E. Farhi},  $9200$ instances were calculated to simulate
the adiabatic algorithm for finding cliques in a random graph in
 Ref. \cite{Childs2}, and $500$ instances were calculated to simulate
the adiabatic algorithm for factorizing integers in Ref.
\cite{Xinhua Peng}. Yet, the approach to illustrate the validity
of an adiabatic algorithm by numerical simulation is restricted by
the numbers of instances as well as by the size of the problem,
i.e., the number of the qubits needed, which has been no more than
$20$ in all these examples. So far, there has not been an effective approach
to identify what kinds of $H(s)$ are always with a nonzero energy gap.

In this paper, we address the validity issue of the adiabatic algorithm.  We present a sufficient condition for the existence of a nonzero energy gap. It can help to effectively identify a large class of the Hamiltonians without energy-level crossing between the ground and excited states, as shown in Fig. 1. As an example of its application, we use it to examine the Hamiltonians used in the previous papers, which show that all these Hamiltonians belong to this class and therefore have a nonzero energy gap.

Based on practical applications, our discussion focuses on the Hamiltonians defined by Eq. (\ref{eq:H(s)}) with Eq. (\ref{problem H1}). To make our results clear, we state our main finding as the following theorem.

\textit{\textbf{Theorem.}}
Let $H(s)=(1-s)H_i+sH_p$ be the time-dependent Hamiltonian of a $d$-dimensional quantum computer, where the problem Hamiltonian $H_p$ is diagonal in the computational basis. If the initial Hamiltonian $H_i$ satisfies in the computational basis:\\
$(1)$ $H_i$ has a unique ground state,
\begin{eqnarray}
\ket{\psi_0}=\left(\begin{array}{c} r_1e^{i\alpha_l} \\r_2e^{i\alpha_2}\\\vdots\\ r_d e^{i\alpha_d}\end{array}\right)\equiv U\left(\begin{array}{c} r_1\\r_2\\\vdots\\ r_d\end{array}\right),\nonumber
\end{eqnarray}
where all $r_i$ are positive numbers satisfying $\sum r_i^2=1$, and
\begin{eqnarray}
U=\textrm{diag}(e^{i\alpha_1},~e^{i\alpha_2},\dots, e^{i\alpha_n})\nonumber
\end{eqnarray}
is a diagonal unitary matrix, and \\
$(2)$ all the nondiagonal elements of $U^\dag H_iU$ are nonpositive,\\
then the energy gap between the ground and excited states of $H(s)$ is nonzero for $s\in[0,1)$.

The theorem indicates that a nonzero energy gap between the ground and excited states of the time-dependent Hamiltonian $H(s)$ is guaranteed if the initial Hamiltonian $H_i$ is  properly chosen such that the two conditions are fulfilled. Despite the fact that these conditions restrict the choice of Hamiltonians, there is actually a large class of  Hamiltonians that satisfy the conditions. For example, all the Hamiltonians for adiabatic algorithms in the previous works, to our knowledge, belong to this class.

We now prove the theorem in three steps.

First, we establish an auxiliary function,
\begin{eqnarray}
F(s)\equiv \left[(1-s)c_1+sc_2\right]I-U^\dag H(s)U, ~~s\in [0,1), \label{A(s)}
\end{eqnarray}
where $c_1$ is a positive number but larger than the largest eigenvalue of $H_i$, and $c_2$ is a positive number but larger than the largest eigenvalue of $H_p$. By substituting $H(s)=(1-s)H_i+sH_p$ into Eq. (\ref{A(s)}) and using the relation $[H_p,U]=0$, $F(s)$ can be recast as
\begin{eqnarray}
F(s)=(1-s)(c_1I-U^\dag H_iU)+s(c_2I-H_p).
\end{eqnarray}
By definition, $H_p$ is diagonal in the computational basis, and the nondiagonal elements of $U^\dag H_iU$ are nonpositive. Besides, it is a general property of a Hermitian matrix that the diagonal elements of it are not larger than its largest eigenvalue. Hence, $c_1I-U^\dag H_iU$,  $c_2I-H_p$, and $F(s)$ are all non-negative matrices. Hereafter, a matrix $A$ is said to be non-negative if each $(A)_{ij} \geq 0$, and it is denoted by $A \geq 0$. Similarly,  $A > 0$  means that $A$ is a positive matrix, i.e., each element of it is positive, and  $A \geq B$ ($A > B$) means that $A_{ij} \geq B_{ij}$ ($A_{ij} > B_{ij}$)  for all the elements.

Second, we show that there exists a positive integer $N_0$ such that $F^{N_0}(s)>0$.  Since both $c_1I-U^\dag H_iU$ and $c_2I-H_p$ are non-negative,  there is
\begin{eqnarray}
F(s)\geq (1-s)(c_1I-U^\dag H_iU)\geq 0. \label{A1}
\end{eqnarray}
From Eq. (\ref{A1}), we have
\begin{eqnarray}
F^N(s)\geq (1-s)^N(c_1I-U^\dag H_iU)^N, \label{A2}
\end{eqnarray}
where $N$ is an arbitrary positive integer. To derive Eq. (\ref{A2}) from Eq. (\ref{A1}), one may consider two non-negative matrices $A$ and $B$ with $A\geq B$. Simple calculations show $A^2=[B+(A-B)]^2=B^2+B(A-B)+(A-B)B+(A-B)^2$. Since $B(A-B)$, $(A-B)B$, and $(A-B)^2$ are all non-negative matrices, one immediately has $A^2\geq B^2$. Similarly , one may obtain $A^N\geq B^N$ for arbitrary integers $N$, i.e., Eq. (\ref{A2}).

To show the existence of $N_0$, we need to consider the limit of $(c_1I-U^\dag H_iU)^N/(c_1-\varepsilon_0)^N$ for $N\rightarrow \infty$, where $\varepsilon_0$ is the ground-state energy of $H_i$. By using the conditions in the theorem, we have
\begin{eqnarray}
&&\lim_{N\rightarrow \infty}
\frac{1}{(c_1-\varepsilon_0)^N}(c_1I-U^\dag H_iU)^N\nonumber\\
&&=\lim_{N\rightarrow \infty}
\sum_m\frac{\left(c_1- \varepsilon_m\right)^N}{(c_1-\varepsilon_0)^N}U^\dag\ket{\psi_m}\bra{\psi_m}U\nonumber\\
&&=U^\dag\ket{\psi_0}\bra{\psi_0}U=\ket{r}\bra{r}>0, \label{limit}
\end{eqnarray}
where $\varepsilon_m$ and $\ket{\psi_m}$ denote the eigenvalues and eigenstates of $H_i$, respectively. Here, we have used $\ket{r}$ to denote the column matrix $(r_1, r_2, \dots,r_d)^T$. Equation (\ref{limit}) implies that
there exists a sufficiently large number $N_0$ such that $(c_1I-U^\dag H_iU)^{N_0}$ is positive, i.e.,
\begin{eqnarray}
(c_1I-U^\dag H_iU)^{N_0}>0.
\label{A3}
\end{eqnarray}
Equations (\ref{A1}), (\ref{A2}), and (\ref{A3}) show that there exists a positive integer $N_0$ such that
\begin{eqnarray}
F^{N_0}(s)>0. \label{fn}
\end{eqnarray}

Third,  we demonstrate that the energy gap between the ground and excited states of $H(s)$ is nonzero for $s\in [0,1)$, with the aid of the above properties of $F(s)$. Note that we have shown $F^{N_0}(s)>0$. According to the Perron-Frobenius theorem that the eigenvector associated with the largest eigenvalue of a positive matrix is unique \cite{Perron}, $F^{N_0}(s)$ has a unique eigenvector associated with its largest eigenvalue. Since $F(s)$ and $F^{N_0}(s)$ share the same spectral structure, $F(s)$ must have a unique eigenvector to its largest eigenvalue too, and so does $UF(s)U^\dag$. From Eq. (\ref{A(s)}), we have $H(s)=[(1-s)c_1+sc_2]I-UF(s)U^\dag$. It implies that $H(s)$ has a unique eigenvector to its smallest eigenvalue when $F(s)$ has a unique eigenvector to $F(s)$'s largest eigenvalue. Therefore, the energy gap between the ground and excited states of $H(s)$ is nonzero during $s\in [0,1)$ as long as the two conditions in the theorem are fulfilled. This completes the proof of the theorem.

The theorem can help to effectively identify a large class of the Hamitonians that have a nonzero energy gap between the ground and excited states. To illustrate the usefulness of the theorem, we now apply it to the models considered in the previous papers \cite{E. Farhi,T. Hogg,F. Gaitan,Gaitan2,Childs2,Ralf,Hofmann,Schaller,Roland,W. van Dam,S. Garnerone,Xinhua Peng,Amin}. Without the need for complicated calculations, we can immediately confirm that all the Hamiltonians used in these papers belong to this class, i.e., they are with a nonzero energy gap. We examine them case by case.

\textit{Case 1.} This case includes the Hamiltonians used in Refs. \cite{E. Farhi,T. Hogg,F. Gaitan,Gaitan2,Amin}, which involve only individual bit rotations. They can be generally expressed as
\begin{eqnarray}
H_i=a_0I+\sum_{i=1}^n a_i\sigma_x^{i},\label{initial H1}
\end{eqnarray}
where $\sigma_x^i$ is the Pauli operator for the $i$th qubit, $a_0$ is a real number, and $a_i$ are negative numbers.  Equation (\ref{initial H1}) is reduced to the Hamiltonian in Refs. \cite{E. Farhi} if $a_0=\sum_{i=1}^n \frac{d_i}{2}$ and $a_i=-\frac{d_i}{2}$, the Hamiltonian in Ref. \cite{T. Hogg} if $a_0=\sum_{i=1}^n \frac{\omega_i}{2}$ and $a_i=-\frac{\omega_i}{2}$, the Hamiltonian in Ref. \cite{Amin} if $a_0=0$ and $a_i=-\Delta_i$, and the Hamiltonian in Ref. \cite{F. Gaitan,Gaitan2} if $a_0=\frac{n}{2}$ and $a_i=-\frac{1}{2}$,  where $d_i$ are positive integers and $\omega_i$,$\Delta_i$ are positive numbers. The unique ground state of Eq. (\ref{initial H1}) is a uniform superposition of the computational basis states,
\begin{eqnarray}
\ket{\psi_0}=\left(\frac{\ket{0}+\ket{1}}{\sqrt{2}}\right)^{\otimes n}.\label{initial_state}
\end{eqnarray}
Comparing Eq. (\ref{initial_state}) with the general expression for $\ket{\psi_0}$ in the theorem, we have $U=I$, $r_i=1/\sqrt{2^n}$, and hence  $UH_iU^\dag=H_i$. In this case, all $r_i$ are positive and  all the nondiagonal elements of $U^\dag H_iU$ are nonpositive. That is, the initial Hamiltonians in Refs. \cite{E. Farhi,T. Hogg,F. Gaitan,Gaitan2,Amin} satisfy the conditions of our theorem, and therefore they are valid to be used for adiabatic computation.

\textit{Case 2.}  This case includes the Hamiltonians used in Refs. \cite{Childs2,Ralf,Hofmann,Schaller}, which involve two-qubit interactions. The Hamiltonian in  \cite{Childs2} reads
\begin{eqnarray}
H_i=-\frac{1}{2}\sum_{i<j}^n(\sigma_x^i\otimes\sigma_x^j+\sigma_y^i\otimes\sigma_y^j),\label{Hb}
\end{eqnarray}
and the Hamiltonians in Refs. \cite{Ralf,Hofmann,Schaller} can be generally expressed as
\begin{eqnarray}
H_i=a_0I+\sum_{i<j}^n a_{ij}(\sigma_x^i\otimes\sigma_x^j+\sigma_y^i\otimes\sigma_y^j+\sigma_z^i\otimes\sigma_z^j),\label{Ha}
\end{eqnarray}
where $a_0$ is a real number, and $a_{ij}$ are nonpositive numbers. Equation (\ref{Ha}) is  reduced to the Hamiltonian in Ref. \cite{Ralf} if $a_0=0$ and $a_{ij}=-\frac{1}{2}M_{ij}$, the Hamiltonian in Ref. \cite{Schaller} if $a_0=0$ and $a_{ij}=-2|f_{ij}|$, and the Hamiltonian in Ref. \cite{Hofmann} if $a_0=\Omega\sum_{i<j}\frac{n_{ij}}{2}$ and $a_{ij}=-\Omega\frac{n_{ij}}{2}$, where $M_{ij}$ and $n_{ij}$ are non-negative integers, $f_{ij}$ are real numbers, and $\Omega>0$. Due to the symmetry of the quantum systems, these Hamiltonians are block diagonal, and therefore the evolutions of the systems are constrained in each subspace identified by their Hamming weight $k$. In the $k$th subspace, the ground state of $H_i$ is
\begin{eqnarray}
\ket{\psi_0}=\left(\begin{array}{c} n\\k \end{array}\right)^{-1/2}\sum_{h(z)=k}\ket{z},\label{second example}
\end{eqnarray}
where $h(z)$ denotes the Hamming weight of $z$. Comparing Eq. (\ref{second example}) with the general expression for $\ket{\psi_0}$ in the theorem, we have $U=I$, $r_i=\left(\begin{array}{c} n\\k \end{array}\right)^{-1/2}$, and hence  $U^\dag H_iU=H_i$. In this case, all $r_i$ are positive and  all the nondiagonal elements of $U^\dag H_iU$ are nonpositive. That is, these initial Hamiltonians satisfy the conditions of our theorem, and therefore they are valid to be used for adiabatic computation \footnote{The authors in Refs. \cite{Ralf,Hofmann,Schaller} also considered another type of Hamiltonians, expressed as  $H_i=a_0I+\sum_{i<j}^na_{ij}(\sigma_x^i\otimes\sigma_x^j+\sigma_y^i\otimes\sigma_y^j)$. Since they did not give an expression of the initial state, we do not examine the existence of the non-zero energy gap for this type of Hamiltonians here.}.

\textit{Case 3.} This case includes the Hamiltonians used in Refs. \cite{Roland,W. van Dam,S. Garnerone}, which involve many-body interactions. They can be expressed as
\begin{eqnarray}
H_i=I-\ket{\psi_0}\bra{\psi_0},\label{Hc}
\end{eqnarray}
where $\ket{\psi_0}=\frac{1}{\sqrt{d}}\Sigma_{i=1}^d\ket{i}$, and $d$ denotes the dimensions of the quantum system.   $\ket{\psi_0}$ is the unique ground state of the Hamiltonian. Clearly, the initial Hamiltonian defined by Eq. (\ref{Hc}) satisfies the conditions of our theorem too.

\textit{Case 4.} The Hamiltonian used in Ref. \cite{Xinhua Peng} reads
\begin{eqnarray}
H_i=g(\sigma_x^1+\sigma_x^2+\cdots+\sigma_x^n),\label{example}
\end{eqnarray}
where $g$ is a positive number. It describes an $n$-qubit system, in which all the qubits interact with the same magnetic field with strength $g$.
The ground state is
\begin{eqnarray}
\ket{\psi_0}=\left(\frac{\ket{0}-\ket{1}}{\sqrt{2}}\right)^{\otimes n}.\label{example1}
\end{eqnarray}
Comparing Eq. (\ref{example1}) with the general expression for $\ket{\psi_0}$ in the theorem, we have $\ket{\psi_0}=U\left(\frac{\ket{0}+\ket{1}}{\sqrt{2}}\right)^{\otimes n}$ with $U=\sigma_z^1\otimes\sigma_z^2\otimes\cdots\otimes\sigma_z^n$, and hence $U^\dag H_iU=-g(\sigma_x^1+\sigma_x^2+\cdots+\sigma_x^n)$. In this case, all $r_i$ are positive and  all the nondiagonal elements of $U^\dag H_iU$ are nonpositive. It satisfies the conditions of our theorem, and therefore the initial Hamiltonian defined by Eq. (\ref{example}) is valid to be used for adiabatic computation. Note that the Hamiltonian defined by Eq. (\ref{example}) can be also considered as an instance of Case $1$ up to a unitary transformation.

So far, we have checked the Hamiltonians from previous work on adiabatic quantum computation and shown that they obey the conditions of our theorem.

\begin{figure}[htbp]
\begin{center}
\includegraphics[width=.25\textwidth]{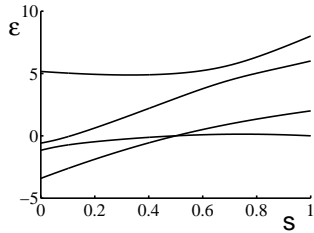}
\end{center}
\caption{Numerical simulation of energy levels for the illustrative example.}
\label{fig.3}
\end{figure}

After having shown that all the Hamiltonians used in the previous works belong to the class identified by our theorem,  we now give an example to illustrate that if the conditions of the theorem are not met, the energy-level crossing between the ground and excited states may happen. Let $H_i=-2\sigma_x\otimes I+I\otimes\sigma_x+I\otimes\sigma_z-2\sigma_x\otimes\sigma_x$ and  $H_p=\textrm{diag}(0,2,6,8)$ in the computational basis. For this example, $\ket{\psi_0}=\frac{\sqrt{4+2\sqrt{2}}}{4}(\sqrt{2}-1,1,\sqrt{2}-1,1)^T$. Comparing it with the general expression in the theorem, we have $ r_i>0$, $U=I$, and hence $U^\dag H_iU=H_i$. It follows that the Hamiltonian fulfills the first condition in the theorem but does not meet the second one. In this case, the Hamiltonian may not be valid for adiabatic quantum computation. Indeed, numerical simulation shows
that the energy-level crossing between the ground and excited states occurs during the evolution time (see Fig. \ref{fig.3}).

Our theorem provides a simple approach to examine the existence of a nonzero energy gap for the Hamiltonian used in the adiabatic quantum algorithm. It may be helpful in choosing alternative Hamiltonians for an adiabatic algorithm. As shown above, all the initial states in the previous papers are an equal-weight superposition of computational bases. However, the theorem shows that it is not necessary for the coefficients $r_i$ to be equal. The theorem only requires that $r_i$ are positive and the nondiagonal elements of $U^\dag H_iU$ are nonpositive, which can sufficiently guarantee the validity of the Hamiltonian being with a nonzero energy gap. This gives a reference scheme for constructing initial Hamiltonians. Indeed, in some instances, the solutions to problems may be more likely to be found in some certain region of state space rather than another. Thus, one may like to input a nonuniform prior distribution into the adiabatic algorithm by following the requirement of the theorem.
Besides, it is also interesting to note that the existence of a nonzero energy gap is independent of the elements of $H_p$ in the computational basis as long as $H_i$ fulfills the conditions in the theorem. In passing, we would like to point out that although the statement of the theorem is based on the expression $H(s)=(1-s)H_i+sH_p$ with $s=t/T$, the theorem is also valid for the general interpolation scheme, $H(t)=a(t)H_i+b(t)H_p$, as long as $a(t)$ and $b(t)$ are monotonic functions satisfying $a(0)=1$, $b(0)=0$, $a(T)=0$, and $b(T)=1$.

Our discussions have focused on the validity issue of the Hamiltonian used in the adiabatic algorithm, i.e.,  the existence of a finite runtime $T$, which is determined by a nonzero energy gap. A Hamiltonian being without level crossing is only a necessary condition for the adiabatic quantum computation. It should be noted that the gap between two noncrossing levels may still be exponentially small or even worse \cite{Farhi2008,Note3}. Another fundamental issue is on the efficiency of the adiabatic algorithm, i.e., the scaling of the runtime, which depends on the value of the nonzero gap. The  efficiency  or the computational complexity of the adiabatic computation is described by the scaling of the inverse square of minimum energy gap, which is related to the problem size \cite{E. Farhi,M. Born}. However, it is quite difficult to theoretically analyze the changing trend of the energy gap for a general Hamiltonian since the dimensions of the time-dependent Hamiltonian increase exponentially with the problem size. We do not attempt to resolve the efficiency issue in this paper.

In conclusion, we put forward a theorem on the existence of a
nonzero energy gap for the Hamiltonians in adiabatic quantum
computation. It can help to effectively identify a large class of
the Hamiltonians with a nonzero energy gap. We have used the theorem to examine the validity of the Hamiltonians in previous papers, and it shows that all the Hamiltonians, examined by us, belong to this class.



\begin{thebibliography}{99}
\bibitem{E. Farhi} E. Farhi, J. Goldstone, S. Gutmann, and M. Siper, arXiv:quant-ph/0001106; E. Farhi, J. Goldstone, S. Gutmann, J. Lapan, A. Lundgren, and D. Preda, Science {\bf 292},
472 (2001).
\bibitem{D. Deutsch} D. Deutsch, Proc. R. Soc. A {\bf 425},
73 (1989).
\bibitem{D. P. Divincenzo} D. P. DiVincenzo, Fort. Phys. {\bf
48}, 771 (2000).
\bibitem{W. van Dam} W. van Dam, M. Mosca, and U. Vazirani,
\textit{Proceedings of the
42nd Symposium on Foundations of Computer Science} (IEEE Computer Society Press, Los Alamitos, 2001), pp. 279--287.
\bibitem{D. Aharonov} D. Aharonov, W. van Dam, J. Kempe, Z. Laudau, S. Lloyd, and O. Regev, SIAM J. Comput. \textbf{37}, 166 (2007).
\bibitem{A. Mizel}  A. Mizel, D. A. Lidar, and M. Mitchell, Phys. Rev.
Lett. {\bf 99}, 070502 (2007).
\bibitem{Childs} A. M. Childs, E. Farhi, and J. Preskill, Phys. Rev.
A \textbf{65}, 012322 (2001).
\bibitem{J. Roland} J. Roland and N. J. Cerf, Phys. Rev. A
\textbf{71}, 032330 (2005).
\bibitem{J.Aberg} J. {\AA}berg, D. Kult, and E. Sj\"{o}qvist, Phys. Rev. A
\textbf{71}, 060312 (2005).
\bibitem{Roland} J. Roland and N. J. Cerf,
Phys. Rev. A {\bf
65}, 042308 (2002).
\bibitem{Childs2} A. M. Childs, E. Farhi, J. Goldstone, and S.
Gutmann, Quantum Inf. Comput. \textbf{2}, 181 (2002).
\bibitem{T. Hogg} T. Hogg, Phys. Rev. A \textbf{67}, 022314 (2003).
\bibitem{Ralf} R. Sch\"{u}tzhold and G. Schaller,
Phys. Rev. A
\textbf{74}, 060304 (2006).
\bibitem{Xinhua Peng} X. Peng, Z. Liao, N. Xu, G. Qin, X. Zhou, D. Suter, and J. Du, Phys. Rev. Lett.
{\bf 101}, 220405 (2008).
\bibitem{Schaller} G. Schaller and R. Sch\"{u}tzhold, Quantum Inf. Comput. \textbf{10}, 0109 (2010).
\bibitem{Amin} N. G. Dickson and M. H. S. Amin, Phys. Rev. Lett. \textbf{106}, 050502 (2011).
\bibitem{F. Gaitan} F. Gaitan and L. Clark, Phys. Rev. Lett. {\bf
108}, 010501 (2012).
\bibitem{S. Garnerone} S. Garnerone, P. Zanardi, and D. A. Lidar,
Phys. Rev. Lett. \textbf{108}, 230506 (2012).
\bibitem{Hofmann} M. Hofmann and G. Schaller, Phys. Rev. A \textbf{89}, 032308 (2014).
\bibitem{Gaitan2} F. Gaitan and L. Clark, Phys. Rev. A \textbf{89}, 022342
(2014).
\bibitem{Steffen} M. Steffen, W. van Dam, T. Hogg, G. Breyta, and I. L. Chuang, Phys. Rev. Lett. {\bf
90}, 067903 (2003).
\bibitem{R.Harris} R. Harris \textit{et al.},
Phys. Rev. B
\textbf{82}, 024511 (2010).
\bibitem{Johnson} M. W. Johnson \textit{et al.}, Nature (London) \textbf{473}, 194 (2011).
\bibitem{Zheng} Z. Bian, F. Chudak, W. G. Macready, L. Clark, and F. Gaitan, Phys. Rev. Lett. \textbf{111}, 130505 (2013).
\bibitem{M. Born} M. Born and V. Fock, Z. Phys. {\bf 51}, 165 (1928).
\bibitem{Bohm}D. Bohm, \textit{Quantum Theory} (Prentic-Hall, New York, 1951).
\bibitem{Messiah}A. Messiah, \textit{Quantum Mechanics} (North-Holland, Amsterdam, 1962).
\bibitem [{Note1()}]{Note1}%
  \BibitemOpen
  \bibinfo {note} {A level crossing at the end of the runtime does not affect the
  validity of the algorithm, since each ground state of $H_p$ encodes a
  solution to the problem.}\BibitemShut {Stop}%
\bibitem{Perron} C. D. Meyer,  \textit{Matrix
Analysis and Applied Linear Algebra} (Society for Industrial and Applied Mathematics, Philadelphia, 2000).
\bibitem [{Note2()}]{Note2}%
  \BibitemOpen
  \bibinfo {note} {The authors in Refs. \cite {Ralf,Hofmann,Schaller} also
  considered another type of Hamiltonian, expressed as $H_i=a_0I+\DOTSB \sum_{i<j}^na_{ij}(\sigma _x^i\otimes \sigma _x^j+\sigma _y^i\otimes
  \sigma _y^j)$. Since they did not give an expression of the initial state, we
  do not examine the existence of the nonzero energy gap for this type of
  Hamiltonian here.}\BibitemShut {Stop}%
\bibitem{Farhi2008} E. Farhi, J. Goldstone, S. Gutmann, and D. Nagaj, Int. J. Quantum Inf. \textbf{6}, 503 (2008).
\bibitem [{Note3()}]{Note3}%
  \BibitemOpen
  \bibinfo {note}
{It is also worth noting that the annealing schedule provided by R. D. Somma, D. Nagaj, and M. Kieferova, Phys. Rev. Lett. \textbf{109}, 050501 (2012), for solving the glued-trees problem, is still efficient even though the minimum energy gap of the Hamiltonian is exponentially small in the problem size.}\BibitemShut {Stop}%
\end{thebibliography}
\end{document}